\documentclass[showpacs,nofootinbib,showkeys,eqsecnum,prd,aps,preprint]{revtex4}

\usepackage[english]{babel}
\usepackage[cp1251]{inputenc}
\usepackage[T1]{fontenc}

\usepackage{amssymb}
\usepackage{amsmath}
\usepackage{amsfonts}
\usepackage{latexsym}
\usepackage{mathrsfs}
\usepackage{bm}
\usepackage{tensor}
\usepackage{hyperref}

\begin{document}

\title{Noncommutative reduction of the nonlinear Schr\"{o}dinger equation
on Lie groups}

\author{A. I. Breev }
\email{breev@mail.tsu.ru}
\affiliation{Department of Theoretical Physics, Tomsk State University Novosobornaya
	Sq. 1, 634050 Tomsk, Russia}

\author{A. V. Shapovalov }
\email{shpv@phys.tsu.ru}
\affiliation{Department of Theoretical Physics, Tomsk State University Novosobornaya
	Sq. 1, 634050 Tomsk, Russia;}
\affiliation{Tomsk Polytechnic University,  Lenin ave., 30, 634034 Tomsk, Russia}

\author{D. M. Gitman }
\email{dmitrygitman@hotmail.com}
\affiliation{Department of Theoretical Physics, Tomsk State University Novosobornaya
	Sq. 1, Tomsk, Russia, 634050;}
\affiliation{P.N. Lebedev Physical Institute, 53 Leninskiy ave., 119991 Moscow, Russia;}
\affiliation{Institute of Physics, University of S\~{a}o Paulo, Rua do Mat\~{a}o, 1371, CEP 05508-090, S\~{a}o Paulo, SP, Brazil.}

\begin{abstract}
We propose a new approach that allows one to reduce nonlinear equations
on Lie groups to equations with a fewer number of independent variables
for finding particular solutions of the nonlinear equations. The main
idea is to apply the method of noncommutative integration to the linear
part of a nonlinear equation, which allows one to find bases in the
space of solutions of linear partial differential equations with a
set of noncommuting symmetry operators. The approach is implemented
for the generalized nonlinear Schr\"{o}dinger equation on a Lie group
in curved space with local cubic nonlinearity. General formalism is
illustrated by the example of noncommutative reduction of the nonstationary
nonlinear Schr\"{o}dinger equation on the motion group $E(2)$ of
the two-dimensional plane $\mathbb{R}^{2}$. In the particular case,
we come to the usual ($1+1$)-dimensional nonlinear Schr\"{o}dinger
equation with the soliton solution. Another example provides the noncommutative
reduction of the stationary multidimensional nonlinear Schr\"{o}dinger
equation on the four-dimensional exponential solvable group.
\end{abstract}



\maketitle

\section{Introduction}

The Lie group theory provides powerful methods for studying linear
and nonlinear differential equations in mathematical physics. Generally,
for the equation with a symmetry group, one can efficiently find and
classify group invariant solutions and conservation laws, generate
new solutions from those already found (see, for example, the well-known
books of Ovsyannikov, Ibragimov, Olver \cite{Ovs,Ibr,Olv}, and many
others).

Remarkable potentialities for finding explicit solutions are opened
up when an equation can be represented directly in terms of the coordinates
of a Lie group. For example, equations on a curved space with a simply
transitive motion group can be represented as equations on a Lie group
manifold. We call such an equation the equation on the Lie group.
Some aspects of integrability of nonlinear equations on Lie groups
are the subject of the present work.

Here, we propose a new approach based on the Lie group theory that
allows one to reduce a nonlinear equation presented in terms of a
Lie group to an equation with a fewer number of independent variables
using the noncommutative ansatz of the work \cite{SpSh1} determined
by the linear part of the nonlinear equation. The noncommutative integration
method (NIM) has been proposed for linear partial differential equations
(PDEs) in \cite{SpSh1}. Following this method, one can find a basis
for the solution space of the linear equation admitting a set of noncommuting
symmetry operators related to the Lie group of invariance of the equation.
Then the noncommutative reduction of a nonlinear equation on a Lie
group yields families of particular solutions containing the parameters
("quantum numbers") of the basis of solutions
to the corresponding linear equation. We describe the proposed noncommutative
reduction for the nonlinear Schr\"{o}dinger equation (NLSE) in curved
space with local cubic nonlinearity and simply transitive motion group
written in terms of the Lie group. The general formalism is illustrated
by the examples of noncommutative reduction of the multidimensional
NLSE on the Lie group $E(2)$ of the two-dimensional plane $\mathbb{R}^{2}$
and on the four-dimensional exponential solvable group. A family of
particular solutions of the NLSE on the Lie group obtained within
the framework of our approch contains the parameters of solutions
of the corresponding linear Schr\"{o}dinger equation.

The nonlinear Schr\"{o}dinger equation is one of the fundamental equations
in nonlinear theoretical physics and mathematics. It arises in a number
of nonlinear models of various physical phenomena and in wide range
of applications. As an example, we recall the theory of optical pulse
propagation in nonlinear media \cite{hasegawa,sulem}. In the theory
of Bose-Einstein condensates, the NLSE is referred to as the Gross-Pitaevskii
equation (GPE) \cite{gpe,PitaevStrin,Keverkid}. The ($1+1$) dimensional
NLSE is integrable within the framework of the soliton theory (see,
e.g., \cite{zakh} and references therein).

The approach proposed here expands the possibilities of constructing
exact solutions of field equations in curved spaces in addition to
the method of separation of variables, which is widely used in general
relativity (see, e.g., recent papers \cite{Mag1,Ob1} and references
therein) and cosmology \cite{El1,Od1,Cap1,Br1}.

We also emphasize that here we consider the noncommutative reduction
of nonlinear equations with \emph{local nonlinearity} in contrast
to the papers \cite{SpSh1,darbu,br2020} where NIM was applied to
equations with nonlocal nonlinearity of convolution type.

The paper is structured as follows. In Section \ref{S2}, we present
the required concepts and definitions from the theory of Lie groups,
introduce notations, and the problem setup. In Section \ref{sec:lambda},
we describe a special representation of the Lie algebra which is constructed
using the orbit method. Then we apply an ansatz for the non-commutative
reduction of the nonlinear Schr\"{o}dinger equation on the Lie group.
Section \ref{sec:ex1} illustrates general approach by the example
of the noncommutative reduction of the nonstationary nonlinear Schr\"{o}dinger
equation (\ref{nsh}) on the motion group $E(2)$ of the two-dimensional
plane $\mathbb{R}^{2}$. In the particular case, we come to the usual
$(1+1)$-dimensional NLSE with the soliton solution. In Section \ref{ex2},
the noncommutative reduction of the stationary multidimensional NLSE
is studied in the case of the four-dimensional exponential solvable
group. In Section \ref{S6}, the concluding remarks are given.

\section{Notations and the problem setup \label{S2}}

In this section, we briefly review the required concepts and definitions
from the Lie group theory and introduce the technical notations.

Let $G$ be an $n-$ dimensional Lie group, its Lie algebra $\mathfrak{g}$
be the tangent space at the group unity $e\in G$, and $\{e_{a}\}$
be a fixed basis in the linear space $\mathfrak{g}$ ($a,b,c=1,\dots,n$).
The Lie group $G$ acts on itself as the left, $L_{\tilde{g}}(g)=\varphi(\tilde{g},g)-$
, and the right, $R_{\tilde{g}}(g)=\varphi(g,\tilde{g})-$, translations,
where $\varphi(g,\tilde{g})$ is a composition function, and $g,\tilde{g}\in G$.
The differentials of the left and right translations determine the
left-invariant, $\xi_{X}(g)=(L_{g})_{*}X$, and the right-invariant,
$\eta_{X}(g)=-(R_{g})_{*}X$, vector fields on the Lie group $G$
($X\in\mathfrak{g}$). Also, we have
\begin{equation}
[\xi_{X},\xi_{Y}]=\xi_{[X,Y]},\quad[\eta_{X},\eta_{Y}]=\eta_{[X,Y]},\quad[\xi_{X},\eta_{Y}]=0,\quad X,Y\in\mathfrak{g},\label{xiEta}
\end{equation}
where $[X,Y]$ is the commutator of $X,Y\in\mathfrak{g}$.

Let $\{e^{b}\}$ be the dual basis to $\{e_{a}\}$ in the Lie algebra
$\mathfrak{g}$, $\langle e^{b},e_{a}\rangle=\delta_{a}^{b}$, and
the brackets $\langle\cdot,\cdot\rangle$ denote the natural pairing
of a 1-form and a vector. Then the left-invariant, $\omega^{X}(g)=(L_{g})^{*}X$,
and the right-invariant, $\sigma^{X}(g)=-(R_{g})^{*}X$, Maurer-Cartan
1-forms satisfy the equations
\begin{equation}
d\omega^{a}=-\frac{1}{2}C_{bc}^{a}\omega^{b}\wedge\omega^{c},\quad d\sigma^{a}=-\frac{1}{2}C_{bc}^{a}\sigma^{b}\wedge\sigma^{c},\quad C_{bc}^{a}=[e_{b},e_{c}]^{a}.\label{CartanEq}
\end{equation}
The implicit summation over repeated indices is assumed.

We take the basis right-invariant vector fields $\eta_{a}(g)=\eta_{e_{a}}(g)$
and the dual right-invariant 1-forms $\sigma^{a}(g)=\sigma^{e^{a}}(g)$
as the moving frame on $G$ and introduce the right-invariant metric
\begin{equation}
ds^{2}=g_{\mu\nu}(g)dg^{\mu}dg^{\nu},\quad\,\mu,\nu=1,\dots,n,\label{ds2}
\end{equation}
where $g^{\mu}$ are local coordinates on $G$. The metric tensor
$g_{\mu\nu}(g)$ of the right-invariant metric (\ref{ds2}) is expanded
over a moving frame with a constant symmetric matrix $G_{ab}$:
\begin{equation}
g_{\mu\nu}(g)=G_{ab}\sigma_{\mu}^{a}(g)\sigma_{\nu}^{b}(g),\quad g^{\mu\nu}(g)=G^{ab}\eta_{a}^{\mu}(g)\eta_{b}^{\nu}(g),\quad\,\,G^{ac}G_{cb}=\delta_{b}^{a}.\label{gGmetric}
\end{equation}

The Christoffel symbols of the symmetric connection consistent with
the metric $ds^{2}$ on the Lie group $G$ are defined in terms of
the metric tensor (\ref{gGmetric}) as
\begin{equation}
\Gamma_{\nu\mu}^{\rho}(g)=\frac{1}{2}g^{\rho\tau}(g)\left(\partial_{\nu}g_{\tau\mu}(g)+\partial_{\mu}g_{\mu\tau}(g)-\partial_{\tau}g_{\nu\mu}(g)\right),\quad\partial_{\nu}\equiv\frac{\partial}{\partial g^{\nu}}.\label{cristG}
\end{equation}

Substituting (\ref{gGmetric}) in (\ref{cristG}) and taking into
account the Maurer-Cartan equations (\ref{CartanEq}), we get (see
Ref. \cite{bar2002}):
\begin{align*}
\Gamma_{\nu\mu}^{\rho}(g) & =\Gamma_{bd}^{a}\sigma_{\nu}^{b}(g)\sigma_{\mu}^{d}(g)\eta_{a}^{\rho}(g)+\eta_{a}^{\rho}(g)\frac{\partial\sigma_{\nu}^{a}(g)}{\partial g^{\mu}},\\
\Gamma_{bd}^{a}(g) & =-\frac{1}{2}C_{bd}^{a}-\frac{1}{2}G^{ac}\left(G_{eb}C_{dc}^{e}+G_{ed}C_{bc}^{e}\right).
\end{align*}
To simplify the presentation, we consider unimodular Lie groups when
the left Haar measure $d\mu_{L}(g)$ coincides with the right Haar
measure on the Lie group $G$: $d\mu_{R}(g)=d\mu_{L}(g)=d\mu(g)$.

Now we can consider differential equations on Lie groups. The Schr\"{o}dinger
equation on a unimodular Lie group $G$ with the metric (\ref{ds2})
for the wave function $\psi=\psi(t,g)$ has the form
\begin{equation}
i\hbar\frac{\partial\psi}{\partial t}=-\frac{\hbar^{2}}{2m}\Delta_{G}\psi,\label{CleinGordon}
\end{equation}
where $\hbar$ is the Planck constant, $m\,(>0)$ is the mass of the
particle, $t$ is the time. The Laplace operator $\Delta_{G}$ on
the Lie group $G$ is a quadratic polynomial in the right-invariant
vector fields:
\begin{equation}
-\hbar^{2}\Delta_{G}=H\left(-i\hbar\eta\right),\quad H(f)=G^{ab}f_{a}f_{b}.\label{dG}
\end{equation}
The operator $\Delta$ is a symmetric operator with respect to the
Riemannian measure
\[
d\mu(g)=\sqrt{\mathrm{det}g_{\mu\nu}}dg=\sqrt{\mathbf{G}}d\mu(g),\quad\mathbf{G}=\det(G_{ab}).
\]

A linear differential operator $X(g)=X(g,\partial_{g})$ commuting
with the operator $H(\eta)$ on some space of functions,
\[
[X(g),H(\eta)]=0,
\]
leaves invariant the set of solutions to the equation and it is the\emph{
symmetry operator} of the equation (\ref{CleinGordon}). From equation
(\ref{xiEta}), one can easily see that the linear equation (\ref{CleinGordon})
admits a set of left-invariant vector fields $\xi_{a}$ as symmetry
operators. It can be shown that the Laplace operator on an $n$-dimensional
manifold admitting a set of $n$ linearly independent symmetry operators
of the first order can always be represented locally in the form (\ref{dG})
up to a constant factor for some Lie group $G$ with right-invariant
metric \cite{brsh2016}.

In this paper, we consider the following nonlinear Schr\"{o}dinger
equation (\ref{CleinGordon}) on the Lie group $G$:
\begin{equation}
i\hbar\frac{\partial\psi}{\partial t}=-\frac{\hbar^{2}}{2m}\Delta_{G}\psi+U\left(g,\psi\right)\psi.\label{nsh}
\end{equation}

Note that the nonlinearity $U(g,\psi)$ does not admit $\xi_{a}$
as symmetry operators of the equation (\ref{nsh}). When $G=\mathbb{R}^{3}$,
we have $U(g,\psi)=\left|\psi\right|^{2}$ and (\ref{nsh}) is the
well-known nonlinear Schr\"{o}dinger equation (see, e.g., \cite{gpe,PitaevStrin,Keverkid,zakh}
, and references therein).

We will show that the NIM is effective for solving the equation (\ref{nsh})
under some restrictions on the Lie group $G$.

\section{Noncommutative reduction of the nonlinear Schr\"{o}dinger equation
\label{sec:lambda}}

The approach to noncommutative reduction of the eqiuation (\ref{nsh})
is based on a special representation of the Lie algebra $\mathfrak{g}$
constructed in terms of the orbit method. We also need a suitable
direct and inverse Fourier transform on the Lie group $G$.

First, we recall some necessary definitions from the orbit method
that will be used hereinafter.

The degenerate Poisson-Lie bracket,
\begin{equation}
\left\{ \phi,\psi\right\} (f)=\langle f,\left[d\phi(f),d\psi(f)\right]\rangle=C_{ab}^{c}f_{c}\frac{\partial\phi(f)}{\partial f_{a}}\frac{\partial\psi(f)}{\partial f_{b}},\quad\phi,\psi\in C^{\infty}(\mathfrak{g}^{*}),\label{pl1-1}
\end{equation}
endows the space $\mathfrak{g}^{*}$ with a Poisson structure \cite{kirr1978}.
Here, $f_{a}$ are the coordinates of a linear functional $f=f_{a}e^{a}\in\mathfrak{g}^{*}$
relative to the dual basis $\left\{ e^{a}\right\} $. The number $\mathrm{ind\mathfrak{g}}$
of functionally independent Casimir functions $K_{\mu}(f)$ with respect
to the bracket (\ref{pl1-1}) is called the index of the Lie algebra
$\mathfrak{g}$, $\mu=1,\dots,\mathrm{ind\mathfrak{g}}.$

A coadjoint representation $\mathrm{Ad}^{*}$: $G\times\mathfrak{g}^{*}\rightarrow\mathfrak{g}^{*}$
splits $\mathfrak{g}^{*}$ into coadjoint orbits (K-orbits). Restriction
of the bracket (\ref{pl1-1}) to an orbit is nondegenerate and coincides
with the Poisson bracket generated by the Kirillov symplectic form
$\omega_{\lambda}$ \cite{kirr1978}. Orbits of maximum dimension
$\mathrm{dim}\mathcal{O}^{(0)}=\mathrm{dim}\mathfrak{g}-\mathrm{ind}\mathfrak{g}$
are called \emph{non-degenerate}. \cite{kirr1978,10_Kirill}.

Let $\mathcal{O}_{\lambda}$ be a nondegenerate K-orbit passing through
the covector $\lambda\in\mathfrak{g}^{*}$. Using Kirillov's orbit
method \cite{10_Kirill}, we construct an unitary irreducible representation
of the Lie group $G$ with respect to a given orbit. This representation
can be constructed iff for the functional $\lambda$ there exists
a subalgebra $\mathfrak{h}\subset\mathfrak{g}^{\mathbb{C}}$ in the
complex extension $\mathfrak{g}^{\mathbb{C}}$ of the Lie algebra
$\mathfrak{g}$ satisfying the conditions:
\begin{equation}
\langle\lambda,[\mathfrak{h},\mathfrak{h}]\rangle=0,\quad\mathrm{dim}\mathfrak{h}=\mathrm{dim}\mathfrak{g}-\frac{1}{2}\mathrm{dim}\mathcal{O}_{\lambda}.\label{defp}
\end{equation}
The subalgebra $\mathfrak{h}$ is called the \emph{polarization} of
the functional $\lambda$. Equation (\ref{defp}) assumes that the
functionals from $\mathfrak{g}^{*}$ can be prolonged to $\mathfrak{g}^{\mathbb{C}}$
by linearity. In this paper, to simplify the presentation, we restrict
ourselves to the case when $\mathfrak{h}$ is the real polarization.

Now we introduce a special coordinate system on the Lie group $G$
compatible with non-degenerate K-orbits of $G$. Let $H$ be a closed
subgroup in a Lie group $G$ , and $\mathfrak{h}$ be the Lie algebra
of $H$. The Lie group acts on the right homogeneous space $Q\simeq G/H:q'=qg$
and defines a principal bundle with the base $Q$, fibers $H$, and
the canonical projection $\pi:G\rightarrow Q$. Choose a basis $\left\{ e_{\overline{\alpha}}\right\} $
in the subalgebra $\mathfrak{h}$ and a basis $\left\{ e'_{\overline{a}}\right\} $
is in the complementary subspace $\mathfrak{m=\mathfrak{h}^{\perp}}$.
In some trivializing neighborhood $V_{0}$ of the unit of the Lie
group $G$, we introduce the local coordinates of the second kind
\begin{align*}
g(q,h) & =\left(e^{h^{\dim\mathfrak{h}}e_{\overline{\dim\mathfrak{h}}}}e^{h^{\dim\mathfrak{h}-1}e_{\overline{\dim\mathfrak{h}-1}}}\dots e^{h^{1}e_{\overline{1}}}\right)\left(e^{q^{\dim Q}e'_{\overline{\dim Q}}}e^{q^{\dim Q-1}e'_{\overline{\dim Q-1}}}\dots e^{q^{1}e'_{\overline{1}}}\right).
\end{align*}
We fix a section $s:Q\rightarrow G$ of the principal bundle of $G$
by the equality
\begin{eqnarray*}
g(q,h) & = & hs(q).
\end{eqnarray*}
The left-invariant vector fields on $G$ in local coordinates $(q,h)$
have the form
\begin{align*}
\xi_{X}(q,h) & =\xi{}_{X}^{\overline{a}}(q)\partial_{q^{\overline{a}}}+\xi_{X}^{\overline{\alpha}}(q,h)\partial_{h^{\overline{\alpha}}},
\end{align*}
where $\alpha_{X}(q)=\xi_{X}^{\overline{a}}(q)\partial_{q^{\overline{a}}}$
are the generators of the group action on the homogeneous space $Q$.

According to the orbit method \cite{kirr1978}, we introduce a unitary
one-dimensional irreducible representation of the Lie group $G$,
which in a neighborhood of $V_{0}$ is given by
\begin{equation}
U^{\lambda}(e^{X})=\exp\left(\frac{i}{\hbar}\langle\lambda,X\rangle\right),\quad X\in\mathfrak{h}.\label{Ul}
\end{equation}
The representation of the Lie group $G$ corresponding to the orbit
$\mathcal{O}_{\lambda}$ is induced by the one-dimensional representation
\begin{align}
 & (T_{g}^{\lambda}\psi)(q)=\Delta_{H}^{-1/2}(h(q,g))U^{\lambda}(h(q,g))\psi(qg)=U^{\lambda+i\hbar\beta}(h(q,g))\psi(qg),\label{TgG}\\
 & \beta_{\overline{\alpha}}=-\frac{1}{2}\mathrm{Tr}\left(\mathrm{\left.ad_{\overline{\alpha}}\right|}_{\mathfrak{h}}\right),\nonumber
\end{align}
where $\Delta_{H}(g)=\mathrm{det}\mathrm{Ad}_{h}$ is the module of
the subgroup $H$, $h\in H;e_{H}$ is the unit element in the Lie
group $H$. The function $h(q,g)$ in (\ref{TgG}) is a factor of
the homogeneous space $Q$:
\[
s(q)g=h(q,g)s(qg),\quad h(q,e)=1.
\]

Let $L_{2}(Q,\mathfrak{h},\lambda)$ denotes the space of functions
defined on $Q$ where the representation (\ref{TgG}) acts. Restriction
of the left-invariant vector fields $\xi_{X}(g)$ on the homogeneous
space $Q$ reads
\begin{align}
 & \ell_{X}(q,\partial_{q},\lambda)=\left.\left(\left[U^{\lambda+i\hbar\beta}(h)\right]^{-1}\xi_{X}(g)U^{\lambda+i\hbar\beta}(h)\right)\right|_{h=e_{H}},\label{getEll-1}\\
 & [\ell_{X}(q,\partial_{q},\lambda),\ell_{Y}(q,\partial_{q},\lambda)]=\ell_{[X,Y]}(q,\partial_{q},\lambda),\quad X,Y\in\mathfrak{g}.\nonumber
\end{align}
The representation (\ref{TgG}) is unitary with respect to the scalar
product in the space of functions $L_{2}(Q,\mathfrak{h},\lambda)$:
\begin{equation}
(\psi_{1},\psi_{2})=\int_{Q}\overline{\psi_{1}(q)}\psi_{2}(q)d\mu(q),\quad d\mu(q)=\rho(q)dq^{1}\dots dq^{l}.\label{scQ}
\end{equation}
The function $\rho(q)$ is determined from the condition that the
operators $-i\ell_{X}(q,\partial_{q},\lambda)$ are Hermitian with
respect to the given scalar product (\ref{scQ}).

The irreducible representation of the Lie algebra $\mathfrak{g}$
by linear operators of the first order (\ref{getEll-1}) depending
on $\mathrm{\dim Q=dim}\mathcal{O}_{\lambda}/2=(\dim\mathfrak{g}-\mathrm{ind}\mathfrak{g})/2$
variables is called the $\lambda$-\emph{representation} of the Lie
algebra $\mathfrak{g}$ . It was introduced in \cite{SpSh1}.

The explicit form of the $\lambda$-representation operators is determined
by left-invariant vector fields in the trivialization domain $V_{0}$
of the principal bundle $G$:
\[
\ell_{X}(q,\partial_{q},\lambda)=\xi_{X}^{\overline{a}}(q)\partial_{q^{\overline{a}}}+\frac{i}{\hbar}\xi_{X}^{\overline{\alpha}}(q,e_{H})\left(\lambda_{\overline{\alpha}}+i\hbar\beta_{\overline{\alpha}}\right).
\]

Let us introduce the direct and inverse generalized Fourier transform,
which is the essential point of the non-commutative integration method.
The representation operators (\ref{TgG}) can be rewritten in the
integral form as
\begin{align}
 & (T_{g}^{\lambda}\psi)(q)=\int_{Q}\psi(q')\mathscr{D}_{qq'}^{\lambda}(g)d\mu(q),\notag\\
 & \mathscr{D}_{qq'}^{\lambda}(g)=\Delta_{H}^{-1/2}(h(q,g))U^{\lambda}(h(q,g))\delta(qg,q'),\notag
\end{align}
where $\delta(q,q')$ is a generalized delta function with respect
to the measure $d\mu(q)$. The generalized kernels $\mathscr{D}_{qq'}^{\lambda}(g)$
of this representation have the properties
\begin{gather*}
\mathscr{D}_{qq'}^{\lambda}(g_{1}g_{2})=\int_{Q}\mathscr{D}_{qq''}^{\lambda}(g_{1})\mathscr{D}_{q''q'}^{\lambda}(g_{2})d\mu(q''),\\
\mathscr{D}_{qq'}^{\lambda}(g)=\overline{\mathscr{D}_{q'q}^{\lambda}(g^{-1})},\quad\mathscr{D}_{qq'}^{\lambda}(e)=\delta(q,q'),
\end{gather*}
where $g_{1},g_{2}\in G$ satisfy the system of equations
\begin{gather}
\left(\eta_{X}(g)+\ell_{X}(q,\partial_{q},\lambda)\right)\mathscr{D}_{qq'}^{\lambda}(g)=0,\quad\left(\xi_{X}(g)+\overline{\ell_{X}(q',\partial_{q'},\lambda)}\right)\mathscr{D}_{qq'}^{\lambda}(g)=0.\label{tD-1}
\end{gather}
Note that the functions $\mathscr{D}_{qq'}^{\lambda}(g)$ are defined
globally on the whole Lie group $G$ if the K-orbit $\mathcal{O}_{\lambda}$
is integer in the sense of Kirillov's definition \cite{10_Kirill}.

The set of generalized functions $\mathscr{D}_{qq'}^{\lambda}(g)$
satisfying the system of equations (\ref{tD-1}) has the properties
of completeness and orthogonality for a certain choice of the measure
$d\mu(\lambda)$ in parameter space $J$:
\begin{gather}
\int_{G}\overline{\mathscr{D}_{\widetilde{q}\widetilde{q}'}^{\widetilde{\lambda}}(g)}\mathscr{D}_{qq'}^{\lambda}(g)d\mu(g)=\delta(q,\tilde{q})\delta(\tilde{q}',q')\delta(\tilde{\lambda},\lambda),\label{Dort-1}\\
\int_{Q\times Q\times J}\overline{\mathscr{D}_{qq'}^{\lambda}(\tilde{g})}\mathscr{D}_{qq'}^{\lambda}(g)d\mu(q)d\mu(\lambda)=\delta(\tilde{g},g),\label{Dful-1}
\end{gather}
where $\delta(g)$ is the generalized Dirac delta function with respect
to the right Haar measure $d\mu(g)$ on the Lie group $G$.

Consider the function space $L(G,d\mu(g))$ of functions of the form
\begin{equation}
\psi(g)=\int_{Q}\psi(q,q',\lambda)\mathscr{D}_{qq'}^{\lambda}\left(g^{-1}\right)d\mu(q')d\mu(q)d\mu(\lambda),\label{psiLD}
\end{equation}
where the function $\psi(q,q',\lambda)$ with respect to the variables
$q$ and $q'$ belongs to the space $L_{2}(Q,\mathfrak{h},\lambda)$.
From (\ref{Dort-1}) and (\ref{Dful-1}), we can write the inverse
transform as
\begin{equation}
\psi(q,q',\lambda)=\int_{Q\times Q\times J}\psi^{\lambda}(g)\overline{\mathscr{D}_{qq'}^{\lambda}\left(g^{-1}\right)}d\mu(g).\label{invF}
\end{equation}
It follows from (\ref{psiLD}) and (\ref{invF}) that the action of
the operators $\xi_{X}(g)$ and $\eta_{X}(g)$ on the function $\psi^{\lambda}(g$)
from $L_{2}(G,\lambda,d\mu(g))$ corresponds to the action of the
operators $\overline{\ell_{X}^{\dagger}(q,\partial_{q},\lambda)}$
and $\ell_{X}(q',\partial_{q'},\lambda)$ on the function $\psi(q,q',\lambda)$:
\begin{align}
 & \xi_{X}(g)\psi^{\lambda}(g)\Longleftrightarrow\overline{\ell_{X}^{\dagger}(q,\partial_{q},\lambda)}\psi(q,q',\lambda)\text{,}\nonumber \\
 & \eta_{X}(g)\psi^{\lambda}(g)\Longleftrightarrow\ell_{X}(q',\partial_{q'},\lambda)\psi(q,q',\lambda).\label{dual}
\end{align}
The functions (\ref{psiLD}) are eigenfunctions for the Casimir operators
$K_{\mu}^{(s)}(i\hbar\xi)=K_{\mu}^{(s)}(-i\hbar\eta)$:
\begin{align*}
 & K_{\mu}^{(s)}(i\hbar\xi)\psi^{\lambda}(g)\Longleftrightarrow\kappa_{\mu}^{(s)}(\lambda)\psi(q,q',\lambda),\\
 & K_{\mu}^{(s)}(-i\hbar\ell(q',\partial_{q'},\lambda))=\kappa_{\mu}^{(s)}(\lambda),\quad\overline{\kappa_{\mu}^{(s)}(\lambda)}=\kappa_{\mu}^{(s)}(\lambda),\quad\lim_{\hbar\rightarrow0}\kappa_{\mu}^{(s)}(\lambda)=\omega_{\mu}^{(s)}(\lambda).
\end{align*}
As a result of the generalized Fourier transform (\ref{psiLD}), the
left and right fields are converted to $\lambda$-representations,
and the Casimir operators become constants.

This fact is core to the method of non-commutative integration of
linear differential equations on Lie groups. The method allows one
to reduce the original linear differential equation
\begin{equation}
-\hbar^{2}\Delta_{G}\psi(g;q,\lambda)=\Lambda^{2}\psi(g;q,\lambda),\quad\Lambda=\mathrm{const}\label{kgf}
\end{equation}
with the number of independent variables $g$ equal to $\dim\mathfrak{g}$
to the equation
\[
H(-i\hbar\ell(q',\partial_{q'},\lambda))\psi(q';q,\lambda)=\Lambda^{2}\psi(q';q,\lambda)
\]
with a fewer number of independent variables $q'$ that is equal to
$(\dim\mathfrak{g}-\mathrm{ind}\mathfrak{g})/2$ using the ansatz
\begin{align}
\psi^{\lambda}(g;q,\lambda) & =U^{\lambda+i\hbar\beta}(h(q,g^{-1}))\psi(qg^{-1};q,\lambda)\label{N-anz-1}
\end{align}
parameterized by $q$ and $\lambda$ . In view of (\ref{Dful-1}),
the set of functions (\ref{N-anz-1}) parameterized by $q$,
$\lambda$ and $\Lambda$ forms a complete set of solutions to the
equation (\ref{kgf}).

Then we will apply the ansatz of the form (\ref{N-anz-1}) to the
non-commutative reduction of the nonlinear Schr\"{o}dinger equation
(\ref{nsh}). Let us look for a solution of (\ref{nsh}) in the form
\[
\psi^{\lambda}(t,g;q)=U^{\lambda+i\hbar\beta}(h(q,g^{-1}))\psi(t,qg^{-1};q,\lambda).
\]

In view of the relations (\ref{dual}), the linear part of the equation
(\ref{nsh}) can be written as
\begin{align*}
 & \left(i\hbar\frac{\partial}{\partial t}+\frac{\hbar^{2}}{2m}\Delta_{G}\right)\psi^{\lambda}(t,g;q)=\\
 & U^{\lambda+i\hbar\beta}(h(q,g^{-1}))\times\\
 & \times\frac{1}{2m}\left.\left[i\hbar\frac{\partial}{\partial t}+H(-i\hbar\ell(q',\partial_{q'},\lambda))\right]\psi(t,q';q,\lambda)\right|_{q'=qg^{-1}},
\end{align*}
and $\left|\psi^{\lambda}(t,g;q)\right|^{2}$ reads

\begin{align*}
\left|\psi^{\lambda}(t,g;q)\right|^{2} & =\left|U^{\lambda+i\hbar\beta}h(q,g^{-1})\right|^{2}\left|\psi(t,qg^{-1};q,\lambda)\right|^{2}=\\
 & =e^{-2h(q,g)\beta_{\overline{\alpha}}}\left|U^{\lambda}h(q,g^{-1})\right|^{2}\left|\psi(t,qg^{-1};q,\lambda)\right|^{2}.
\end{align*}
For the real polarization $\mathfrak{h}$, in view of the formula
(\ref{Ul}), $|U^{\lambda}(h(q,g^{-1}))|=1$. Then, we have
\[
\left|\psi^{\lambda}(t,g;q)\right|^{2}=e^{-2h(q,g)\beta_{\overline{\alpha}}}\left|\psi(t,qg^{-1};q,\lambda)\right|^{2}.
\]

We only consider the Lie groups $G$ for which
\begin{equation}
e^{-2h(q,g)\beta_{\overline{\alpha}}}=\kappa^{2}(q).\label{cond1}
\end{equation}

The condition (\ref{cond1}) is satisfied if the covector $\beta$
is zero. Thus, under the condition (\ref{cond1}), we obtain the reduced
nonlinear Schr\"{o}dinger equation
\begin{align*}
 & \left[i\hbar\frac{\partial}{\partial t}+\frac{1}{2m}H(-i\hbar\ell(q',\partial_{q'},\lambda))\right]\psi(t,q';q,\lambda)+\\
 & +U\left(\kappa^{2}(q)\left|\psi(t,q';q,\lambda)\right|^{2}\right)\psi(t,q';q,\lambda)=0
\end{align*}
with the fewer number of independent variables $q'$.

\section{The three-dimensional group $E(2)$\label{sec:ex1}}

Here, we consider an example of non-commutative reduction of the nonlinear
Schr\"{o}dinger equation (\ref{nsh}) on the motion group $E(2)$
of the two-dimensional plane $\mathbb{R}^{2}$. The three-dimensional
Lie algebra $\mathfrak{e}(2)$ of $E(2)$ is determined by the commutation
relations $[e_{1},e_{3}]=-e_{2}$, $[e_{2},e_{3}]=e_{1}$ relative
to the fixed basis $\{e_{1},e_{2},e_{3}\}$.

The left-invariant and the right-invariant vector fields on a$E(2)$
have the form
\begin{align*}
 & \xi_{1}=\partial_{x},\quad\xi_{2}=\partial_{y},\quad\xi_{3}=y\partial_{x}-x\partial_{y}+\partial_{\alpha},\\
 & \eta_{1}=-\cos\alpha\partial_{x}+\sin\alpha\partial_{y},\\
 & \eta_{2}=-\sin\alpha\partial_{x}-\cos\alpha\partial_{y},\quad\eta_{3}=-\partial_{\alpha}
\end{align*}
with respect to the canonical coordinates $(x,y,\alpha)$ of the second
kind:
\[
g=(x,y,\alpha)=e^{\alpha e_{3}}e^{ye_{1}}e^{xe_{1}},\quad(x,y)\in\mathbb{R}^{2},\quad\alpha\in[0,2\pi).
\]

The invariant measure on the group coincides with the Lebesgue measure
$d\mu(g)=dxdyd\alpha$. The composition law of the group is
\begin{align*}
 & g_{1}g_{2}=\left(x_{2}+x_{1}\cos\alpha_{2}+y_{1}\sin\alpha_{2},y_{2}-x_{1}\sin\alpha_{2}+y_{1}\cos\alpha_{2},\alpha_{1}+\alpha_{2}\right),\\
 & g_{1}=(x_{1},y_{1},\alpha_{1}),\quad g_{2}=(x_{2},y_{2},\alpha_{2}).
\end{align*}

Each non-degenerate orbit is determined by the Casimir function $K(f)=f_{1}^{2}+f_{2}^{2}$
on the dual space $\mathfrak{e}^{*}(2)\simeq\mathbb{R}^{2}$ and passes
through the covector $\lambda(j)=(j,0,0)$, $j>0,$i.e.
\begin{align*}
 & \mathcal{O}_{j}=\{f\in\mathbb{R}^{3}\mid K(f)=j^{2},\neg(f_{1}=f_{2}=0)\},\\
 & \dim\mathcal{O}_{j}=2.
\end{align*}

The $\lambda$-representation operators corresponding to the real
polarization $\mathfrak{h}=\{e_{1},e_{2}\}$ have the form
\[
\ell_{1}=i\frac{j}{\hbar}\cos q,\quad\ell_{2}=-i\frac{j}{\hbar}\sin q,\quad\ell_{3}=\partial_{q},\quad q\in[0;2\pi).
\]
The operators $-i\hbar\ell_{a}$ are symmetric with respect to the
measure $d\mu(q)=dq$, and all non-degenerate orbits are integer.
Solving the system of equations (\ref{tD-1}), we find the functions
$\mathscr{D}_{qq'}^{\lambda}(g^{-1})$, and the completeness and orthogonality
conditions for them yield the following measure $d\mu(\lambda)$:
\begin{align*}
 & \mathscr{D}_{qq'}^{\lambda}(g^{-1})=\exp\left[\frac{ij_{1}}{\hbar}\left(y\sin q-x\cos q\right)\right]\delta\left(q'-q+\alpha\right),\\
 & d\mu(\lambda)=\frac{1}{(2\pi)^{2}}jdj.
\end{align*}

Let us introduce the right-invariant metric given by the matrix $(G^{ab})=\mathrm{diag}(\delta_{1},\delta_{2},\delta_{3})$.
In local coordinates, this metric can be written as
\begin{align}
ds^{2} & =\left(\delta_{1}^{-1}\cos^{2}\alpha+\delta_{2}^{-1}\sin^{2}\alpha\right)dx^{2}+\nonumber \\
 & +\left(\delta_{1}^{-1}\sin^{2}\alpha+\delta_{2}^{-1}\cos^{2}\alpha\right)dy^{2}+\delta_{3}^{-1}d\alpha^{2}.\label{ex1:ds2}
\end{align}
 The metric (\ref{ex1:ds2}) has the nonzero scalar curvature $R=\delta_{3}(\delta_{1}-\delta_{2})^{2}/(2\delta_{1}\delta_{2}),$
and the corresponding Laplace operator reads
\begin{align*}
\Delta_{E(2)} & =\left(\delta_{1}\cos^{2}\alpha+\delta_{2}\sin^{2}\alpha\right)\partial_{xx}^{2}+\\
 & \left(\delta_{1}\sin^{2}\alpha+\delta_{2}\cos^{2}\alpha\right)\partial_{yy}^{2}+(\delta_{2}-\delta_{1})\sin2\alpha\partial_{xy}^{2}+\delta_{3}\partial_{\alpha\alpha}^{2}.
\end{align*}

For the nonlinear Schr\"{o}dinger equation with the Laplace operator
$\Delta_{E(2)}$ and potential $V=V(\alpha)$,
\begin{equation}
i\hbar\frac{\partial\psi}{\partial t}=\left(-\frac{\hbar^{2}}{2m}\Delta_{E(2)}+V(\alpha)-\varepsilon\left|\psi\right|^{2}\right)\psi,\label{ex1:nsh}
\end{equation}
\[
\,\psi(t,g;q,j)=\exp\left[\frac{ij}{\hbar}\left(y\sin q-x\cos q\right)\right]\psi(t,q-\alpha).
\]
Then, for the function $\psi(t,q')$, the equation (\ref{ex1:nsh})
yields the following reduced equation:
\begin{align}
 & i\hbar\frac{\partial\psi(t,q')}{\partial t}+\frac{\hbar^{2}}{2m}\delta_{3}\frac{\partial^{2}\psi(t,q')}{\partial q'^{2}}-\nonumber \\
 & -\left[\frac{\hbar^{2}j}{2m}\text{\ensuremath{\left(\delta_{1}\cos^{2}q'+\delta_{2}\sin^{2}q'\right)+V(q-q')}}-\varepsilon\left|\psi(t,q')\right|^{2}\right]\psi(t,q')=0.\label{ex1:rd}
\end{align}

It can be seen that in the particular case $V(\alpha)=0$, $\delta_{1}=\delta_{2}$,
$\delta_{3}=1$ the equation (\ref{ex1:rd}) takes the form of the
usual nonlinear Schr\"{o}dinger equation and has the soliton solution
\begin{align*}
 & \psi(t,q')=\frac{\hbar a}{\sqrt{\varepsilon m}}\cosh^{-1}\left[\left(q'-vt\right)\right]\exp\left[\frac{im}{\hbar}\left(q'-\frac{v}{2}\right)v-\frac{i\hbar}{2m}\left(a^{2}-\delta_{1}n'^{2}\right)\right],\\
 & j=\hbar n',\quad\varepsilon>0.
\end{align*}
The solution to the original equation (\ref{ex1:nsh}) has the form
\begin{align*}
\psi(t,g;q,n') & =\frac{\hbar a}{\sqrt{\varepsilon m}}\cosh^{-1}\left[\left(q-\alpha-vt\right)\right]\times\\
 & \times\exp\left[i\left(y\sin q-x\cos q\right)n'+\frac{im}{\hbar}\left(q-\alpha-\frac{v}{2}\right)v-\frac{i\hbar}{2m}\left(a^{2}-\delta_{1}n'^{2}\right)\right].
\end{align*}

Concluding this section, we note that the nonlinear equation (\ref{nsh})
on the Lie groups includes as a particular case the well-known classical
($1+1$)-dimensional nonlinear Schr\"{o}dinger equation
integrable by the Inverse Scattering Transform method (e.g., \cite{zakh}),
and the noncommutative reduction method proposed in this paper yields
the one-soliton solution. This case follows from the more general
equation (\ref{ex1:nsh}) with a potential $V(\alpha)$, which can
be regarded as an example of the Gross-Pitaevskii equation \cite{gpe}.

\section{The four-dimensional solvable exponential group\label{ex2}}

Consider a four-dimensional solvable exponential group $G$. The Lie
algebra$\mathfrak{g}$ of $G$, with respect to a fixed basis $\{e_{1},e_{2},e_{3},e_{4}\}$,
is defined by the commutation relations $[e_{2},e_{3}]=e_{1}$, $[e_{2},e_{4}]=e_{2}$,
$[e_{3},e_{4}]=-e_{3}$. The algebra index equals 2 and there are
two Casimir functions
\[
K_{1}(f)=f_{1},\quad K_{2}(f)=f_{1}f_{4}-f_{3}f_{2},\quad f\in\mathfrak{g}^{*}\simeq\mathbb{R}^{4}.
\]

In canonical coordinates of the second kind
\[
g(x_{1},x_{2},x_{3},x_{4})=e^{x_{4}e_{4}}e^{x_{3}e_{3}}e^{x_{2}e_{2}}e^{x_{1}e_{1}},\quad x_{1}\in[0,2\pi),(x_{2},x_{3},x_{4})\in\mathbb{R}^{3},
\]
the left-invariant and the right-invariant vector fields are given
by
\begin{gather*}
\xi_{1}=\partial_{x_{1}},\quad\xi_{2}=\partial_{x_{3}},\quad\xi_{3}=x_{2}\partial_{x_{1}}+\partial_{x_{3}},\quad\xi_{4}=x_{2}\partial_{x_{2}}-x_{3}\partial_{x_{3}},\\
\eta_{1}=\partial_{x_{1}},\quad\eta_{2}=-e^{x_{4}}\left(x_{3}\partial_{x_{1}}+\partial_{x_{2}}\right)\quad\eta_{3}=-e^{-x_{4}}\partial_{x_{3}},\quad\eta_{4}=-\partial_{x_{4}}.
\end{gather*}
The invariant measure on the group coincides with the Lebesgue measure
and is of the form $d\mu(g)=dx_{1}dx_{2}dx_{3}dx_{4}$. The subgroup
$G_{1}=\{\exp(e_{1}x_{1})\}$ of the Lie group $G$ can be either
compact $(x_{1}\ \in[0;2\pi))$ or noncompact $(x_{1}\in\mathbb{R}^{1})$.
Let us choose the right-invariant metric on the group as follows:
\begin{align}
 & ds^{2}=\delta_{1}^{-1}dx_{1}dx_{4}+\left(\delta_{2}^{-1}dx_{3}-\delta_{1}^{-1}x_{3}dx_{4}\right)dx_{2},\label{ex:ds}\\
 & \left(g^{ab}\right)=2\mathrm{antidiag}\left(\delta_{1},\delta_{2},\delta_{2},\delta_{1}\right),\nonumber \\
 & \delta_{2}\neq-\delta_{1},\quad\delta_{1},\delta_{2}=\mathrm{const}.\nonumber
\end{align}
The metric (\ref{ex:ds}) is not flat since there is a nonzero component
of the Ricci tensor $R_{\mu\nu}(g):R_{44}(g)=(\delta_{2}/\delta_{1})^{2}/2$.
The Laplace operator of the metric (\ref{ex:ds}) reads
\[
\Delta_{G}=4\delta_{1}\partial_{x_{1}x_{4}}^{2}+2\delta_{2}\text{\ensuremath{\left(2\partial_{x_{2}x_{3}}^{2}+2x_{3}\partial_{x_{1}x_{3}}^{2}+\partial_{x_{1}}\right)}}.
\]

In this section, we will consider a stationary nonlinear Schr\"{o}dinger
equation of the form
\begin{equation}
-\frac{\hbar^{2}}{2m}\Delta_{G}\psi(g)+\varepsilon e^{x_{4}}\left|\psi(g)\right|^{2}\psi(g)=E\psi(g),\quad E>0.\label{ex:nsh}
\end{equation}
There is a complete set of commuting symmetry operators $\{-i\hbar\xi_{1},-i\hbar\xi_{2},K_{2}(-i\hbar\xi)\}$
that allows one to perform a complete separation of variables in the
linear equation (\ref{ex:nsh}) with $\varepsilon=0$:
\begin{equation}
\psi_{p_{1}p_{2}j_{2}}(g)=e^{\frac{i}{\hbar}\left(p_{1}x_{1}+p_{2}x_{2}\right)}\left(\frac{\hbar}{p_{2}+p_{1}x_{3}}\right){}^{\frac{1}{2}+\frac{ij_{2}}{\hbar p_{1}}}\varphi_{p_{1}p_{2}j_{2}}\left(x_{4}+\ln\frac{p_{2}+p_{1}x_{3}}{\hbar}\right),\label{p1p2j}
\end{equation}
\begin{align*}
 & -i\hbar\xi_{1}\psi_{p_{1}p_{2}j_{2}}(g)=p_{1}\psi_{p_{1}p_{2}j_{2}}(g),\\
 & -i\hbar\xi_{2}\psi_{p_{1}p_{2}j_{2}}(g)=p_{2}\psi_{p_{1}p_{2}j_{2}}(g),\\
 & K_{2}(-i\hbar\xi)\psi_{p_{1}p_{2}j_{2}}(g)=j_{2}\psi_{p_{1}p_{2}j_{2}}(g).
\end{align*}
Substituting the ansatz (\ref{p1p2j}) into the equation (\ref{ex:nsh})
with $\varepsilon=0$, we get the ordinary differential equation
\[
2(\delta_{1}+\delta_{2})p_{1}\frac{d\varphi_{p_{1}p_{2}j_{2}}(z)}{dz}-\frac{i}{\hbar}\left(2\delta_{2}j_{2}+mE\right)\varphi_{p_{1}p_{2}j_{2}}(z)=0.
\]
Nevertheless, it is not possible to reduce the nonlinear equation
(\ref{ex:nsh}) (when $\varepsilon\neq0$) since
\[
e^{x_{4}}\left|\psi_{p_{1}p_{2}j_{2}}(g)\right|^{2}\psi_{p_{1}p_{2}j_{2}}(g)=\frac{e^{z}}{\left(p_{2}+p_{1}x_{3}\right)^{2}}\left|\varphi_{p_{1}p_{2}j_{2}}(z)\right|^{2}\varphi_{p_{1}p_{2}j_{2}}(z)
\]
and the expression $e^{z}/\left(p_{2}+p_{1}x_{3}\right)^{2}$ depends
on the variable $x_{3}$.

Let us now carry out the non-commutative reduction. Each nondegenerate
K-orbit passes through the parameterized covector $\lambda(j)=(j_{1},0,0,j_{2}),j=(j_{1},j_{2})\in\mathbb{R}^{2}$:
\begin{align*}
 & \mathcal{O}_{j}=\{f\in\mathbb{R}^{4}\mid K(f)=j_{1},K(f)=j_{1}j_{2},\neg(f_{1}=f_{2}=f=0)\},\\
 & \dim\mathcal{O}_{j}=2.
\end{align*}

The $\lambda$-representation operators corresponding to nondegenerate
K-orbits and real polarization $\mathfrak{h}=\{e_{1},e_{3},e_{4}\}$
have the form
\begin{align*}
 & \ell_{1}=i\frac{j_{1}}{\hbar},\quad\ell_{2}=\partial_{q},\quad\ell_{3}=i\frac{j_{1}}{\hbar}q,\quad\ell_{4}=q\partial_{q}+\frac{i}{\hbar}\left(j_{2}-i\hbar\frac{1}{2}\right),\\
 & K_{1}(-i\hbar\ell)=j_{1},\quad K_{2}(-i\hbar\ell)=j_{1}j_{2},
\end{align*}
where the covector $\beta=(0,0,0,-1/2)$. The operators $-i\hbar\ell_{a}$
are symmetric with respect to the measure $d\mu(q)=dq,q\in Q\simeq\mathbb{R}^{1}$.

Solving the system of equations (\ref{tD-1}), we obtain the functions
$\mathscr{D}_{qq'}^{\lambda}(g^{-1})$, and the completeness and orthogonality
conditions for them yield the following measute $d\mu(\lambda)$ :
\begin{align*}
 & \mathscr{D}_{qq'}^{\lambda}(g^{-1})=\exp\left(-\frac{1}{2}x_{4}-\frac{ij_{1}}{\hbar}\left(x_{3}\left(q-x_{2}\right)+x_{1}\right)-\frac{ij_{2}}{\hbar}x_{4}\right)\delta\left(q'+e^{-x_{4}}\left(x_{2}-q\right)\right),\\
 & d\mu(\lambda)=\frac{1}{(2\pi)^{3}}j_{1}dj_{1}dj_{2}.
\end{align*}
Then, the non-commutative ansatz has the form
\begin{align}
\psi(g;q,j_{1},j_{2}) & =e^{-x_{4}/2}\exp\left(-\frac{ij_{1}}{\hbar}\left(x_{3}\left(q-x_{2}\right)+x_{1}\right)-\frac{ij_{2}}{\hbar}x_{4}\right)\times\label{ex:anz}\\
 & \times\psi\left(e^{-x_{4}}\left(q-x_{2}\right)\right).\nonumber
\end{align}
Substituting (\ref{ex:anz}) into (\ref{ex1:nsh}), we obtain the
ordinary differential equation
\begin{align}
 & -\frac{n_{1}\hbar^{2}}{m}\left[i(\delta_{1}+\delta_{2})\left(2q'\frac{d}{dq'}+1\right)-2\hbar\delta_{2}n_{2}\right]\psi(q')+\nonumber \\
 & +\varepsilon\left|\psi(q')\right|^{2}\psi(q')=E\psi(q').\label{eq:redeq}
\end{align}
In the linear case $\varepsilon=0$, we have a solution
\[
\psi(q')=\frac{1}{\sqrt{q'}}\exp\left(i\frac{mE/(2\hbar^{2})-\delta_{1}n_{1}\mathit{n}_{2}}{\left(\delta_{1}+\delta_{2}\right)\mathit{n}_{1}}\ln q'\right),\quad\varepsilon=0.
\]

We seek a solution of the equation (\ref{eq:redeq}) in the form
\begin{equation}
\psi(q')=f(q')\exp\left(i\Phi(q')\right),\label{eq:anz22}
\end{equation}
where $f(q')$ and $\Phi(q')$ are real functions. Substituting (\ref{eq:anz22})
in (\ref{eq:redeq}) , we get the ODE system:
\begin{align*}
 & 2\frac{\hbar^{2}}{m}\left(\delta_{1}+\delta_{2}\right)\mathit{j}_{1}q'f'(q')+\varepsilon f(q')^{3}\cot\phi(q')+\\
 & \left[2\frac{\hbar^{2}}{m}\left(\delta_{1}+\delta_{2}\right)\mathit{j}_{1}\left(2q'\phi'(q')\cot\phi(q')+1\right)-E\cot\phi(q')\right]f(q')=0,\\
 & 2q'f'(q')+f(q')=0.
\end{align*}
The solution of this system yields
\begin{equation}
\psi(q')=\sqrt{\frac{\hbar^{2}}{\varepsilon m}\frac{2\left(\delta_{1}+\delta_{2}\right)n_{1}}{q'}}\exp\left\{ i\left[\frac{c_{1}}{q'}+\frac{mE/(2n_{1}\hbar^{2})-\delta_{1}\mathit{n}_{2}}{\delta_{1}+\delta_{2}}\right]\ln q'+\frac{\ln c_{1}}{2}\right\} .\label{ex:sr}
\end{equation}

Substituting (\ref{ex:sr}) into the expression (\ref{ex:anz}), we
obtain a set of particular solutions $\psi(g)$ of the nonlinear equation
(\ref{ex1:nsh}) that are parameterized by $\{q',n_{1},n_{2}\}$ and
$c_{1}$. For this set of solutions, the following equality holds:
\begin{equation}
\left|\psi(g)\right|^{2}=\frac{2\hbar^{2}}{\varepsilon m}\left(\delta_{1}+\delta_{2}\right)\left|\frac{\mathit{n}_{1}c_{1}}{q-x_{2}}\right|.\label{ex:norm}
\end{equation}

Thus, the non-commutative reduction of the equation (\ref{ex1:nsh})
to (\ref{eq:redeq}) made it possible to find a family of particular
solutions of the original equation (\ref{ex1:nsh}). The solutions
obtained tend to infinity on the plane $x_{2}=q$ and tend to zero
as $x_{2}\rightarrow\pm\infty$ that can be seen from (\ref{ex:norm}).

\section*{Conclusion\label{S6}}

In this article, we consider an approach in which the noncommutative
integration method developed in \cite{SpSh1} for finding bases for
solution spaces of linear PDEs with symmetries can be applied to constructing
families of particular solutions of nonlinear equations on Lie groups
by reducing the nonlinear equation to an equation with a fewer number
of independent variables. In terms of this approach, we study the
generalized nonlinear Schr\"{o}dinger equation in curved space with
local cubic nonlinearity on a Lie group.

The application of the noncommutative integration method to nonlinear
equations on Lie groups, under certain restrictions on the Lie group,
allows finding families of particular solutions parameterized by the
eigenvalues of the non-commutative set of symmetry operators for the
\emph{linear part} of the nonlinear equation under consideration.
The nonlinear term in the original nonlinear equation does not admit
those symmetry operators that its linear part admits. On the other
hand, the noncommutative ansatz is determined only by the algebra
of symmetry operators of the linear part of the nonlinear equation.
The special form of the ansatz (\ref{N-anz-1}) and its algebraic
properties allow us in a number of cases to carry out a non-commutative
reduction of the original nonlinear equation.

The parameters $q$ and $\lambda$ in the noncommutative ansatz (\ref{N-anz-1})
acquire a physical meaning when comparing the solution of a nonlinear
equation with the solution of its linear counterpart as it was considered
in \cite{malomed}.

In some cases, it is possible to carry out the noncommutative reduction
to a nonlinear equation with an external potential. In the case of
the NLSE with a potential, we arrive at the Gross-Pitaevskii equation,
which is the model mean field equation in the BEC theory \cite{gpe,PitaevStrin,Keverkid}.
This case is demonstrated by the example of the NLSE with the external
potential (\ref{ex1:nsh}) on the three-dimensional Lie group $E(2)$
in Section \ref{sec:ex1}. With a special choice of the right-invariant
metric on the group $E(2)$, we have obtained the classical ($1+1$)
dimensional NLSE as a result of noncommutative reduction. This made
it possible to obtain a soliton type solution for the NLSE on the
group $E(2)$.

We also note that in this paper we consider the NLSE with local nonlinearity
in contrast to papers \cite{br14,br13}, where the noncommutative
reduction was applied to nonlinear equations with a nonlocal term
of the convolution type. In those papers, the original nonlocal nonlinear
equation was reduced to a nonlocal nonlinear equation with a fewer
number of independent variables using the generalized Fourier transform.

The broad implication of the present research is that the noncommutative
reduction of the NLSE considered in this paper expands the possibilities
of exact integration of nonlinear equations on Lie groups, and, what
is important, in the multidimensional case. The proposed approach
is rather limited by the symmetries of the equation than by its specific
form. Therefore, the proposed version of noncommutative reduction
can be applied to other equations, among which the nonlinear relativistic
equations are of particular interest, for example, the nonlinear Dirac
equation, the sine-Gordon equation, and the reaction-diffusion type
equations. In addition, the problem of the search of nonlinear equations
admitting a noncommutative reduction naturally arises.

\section{Acknowledgments}

The work is supported by Russian Science Foundation, grant No. 19-12-00042.

\end{document}